\documentclass[10pt,conference]{IEEEtran}
\IEEEoverridecommandlockouts
\usepackage{cite}
\usepackage{amsmath,amssymb,amsfonts}
\usepackage{algorithmic}
\usepackage{graphicx}
\usepackage{textcomp}
\usepackage{xcolor}
\def\BibTeX{{\rm B\kern-.05em{\sc i\kern-.025em b}\kern-.08em
    T\kern-.1667em\lower.7ex\hbox{E}\kern-.125emX}}

\usepackage[export]{adjustbox} 
\usepackage{amssymb,amsmath}

\usepackage[bookmarks=false]{hyperref}

\usepackage{listings}

\definecolor{light-gray}{gray}{0.97}
\definecolor{gray}{rgb}{0.4,0.4,0.4}
\definecolor{darkblue}{rgb}{0.0,0.0,0.6}
\definecolor{cyan}{rgb}{0.0,0.6,0.6}

\lstnewenvironment{sql} {\lstset{
	morekeywords={SELECT, FROM, WHERE, ORDER, BY, ASC, LIKE, INNER, JOIN, ON} 
}}{}

\begin{document}

\title{SOTorrent: Studying the Origin, Evolution, and Usage of Stack Overflow Code Snippets}

\author{
\IEEEauthorblockN{Sebastian Baltes}
\IEEEauthorblockA{\textit{University of Trier, Germany} \\
research@sbaltes.com}
\and
\IEEEauthorblockN{Christoph Treude}
\IEEEauthorblockA{\textit{University of Adelaide, Australia} \\
christoph.treude@adelaide.edu.au}
\and
\IEEEauthorblockN{Stephan Diehl}
\IEEEauthorblockA{\textit{University of Trier, Germany} \\
diehl@uni-trier.de}
}

\maketitle

\begin{abstract}
Stack Overflow (SO) is the most popular question-and-answer website for software developers, providing a large amount of copyable code snippets.
Like other software artifacts, code on SO evolves over time, for example when bugs are fixed or APIs are updated to the most recent version.
To be able to analyze how code and the surrounding text on SO evolves, we built \emph{SOTorrent}, an open dataset based on the official SO data dump.
\emph{SOTorrent} provides access to the version history of SO content at the level of whole posts and individual text and code blocks.
It connects code snippets from SO posts to other platforms by aggregating URLs from surrounding text blocks and comments, and by collecting references from GitHub files to SO posts.
Our vision is that researchers will use \emph{SOTorrent} to investigate and understand the evolution and maintenance of code on SO and its relation to other platforms such as GitHub.
\end{abstract}


\section{Introduction}

Stack Overflow (SO) is the most popular question-and-answer website for software developers.
Many of its over 40 million posts~\cite{StackExchangeInc2017d} contain code snippets together with explanations~\cite{YangHussainOthers2016}.
Those snippets do not exist in isolation but are actively reused by developers in their software projects, regardless of maintainability, security, and licensing implications~\cite{BaltesKieferOthers2017, AnMloukiOthers2017, YangMartinsOthers2017,  GharehyazieRayOthers2017, AbdalkareemShihabOthers2017, XiaBaoOthers2017, FischerBottingerOthers2017, AcarBackesOthers2016}.
Yet, there is still a lack of knowledge on how exactly SO code snippets are sourced from and reused on other platforms.
Understanding the evolution of SO content, together with its origin and usage in software projects, is crucial to identify outdated information, detect compatibility issues, and prevent copy-and-paste bugs.

Similar to other software artifacts such as source code files and documentation~\cite{Lehman1980, ChapinHaleOthers2001, MensDemeyer2008, GodfreyGerman2008}, text and code on SO evolve over time, e.g., when the SO community fixes bugs in code snippets or updates documentation to match new API versions.
Since the inception of SO in 2008, a total of 14.8 million SO posts have been edited after their creation---21,601 of them more than ten times.
While many SO posts contain code, the evolution of code snippets on SO differs from the evolution of entire software projects.
Most snippets are relatively short~\cite{BaltesDumaniOthers2018} and many of them cannot compile without modification~\cite{YangHussainOthers2016}.
In addition, SO does not provide a version control or bug tracking system for code snippets, forcing users to rely on the commenting function or additional answers to voice concerns about a snippet.

\section{SOTorrent: MSR mining challenge 2019}

\begin{figure}[h!]
\centering
\includegraphics[width=1\columnwidth,  trim=0.0in 0.0in 0.0in 0.0in]{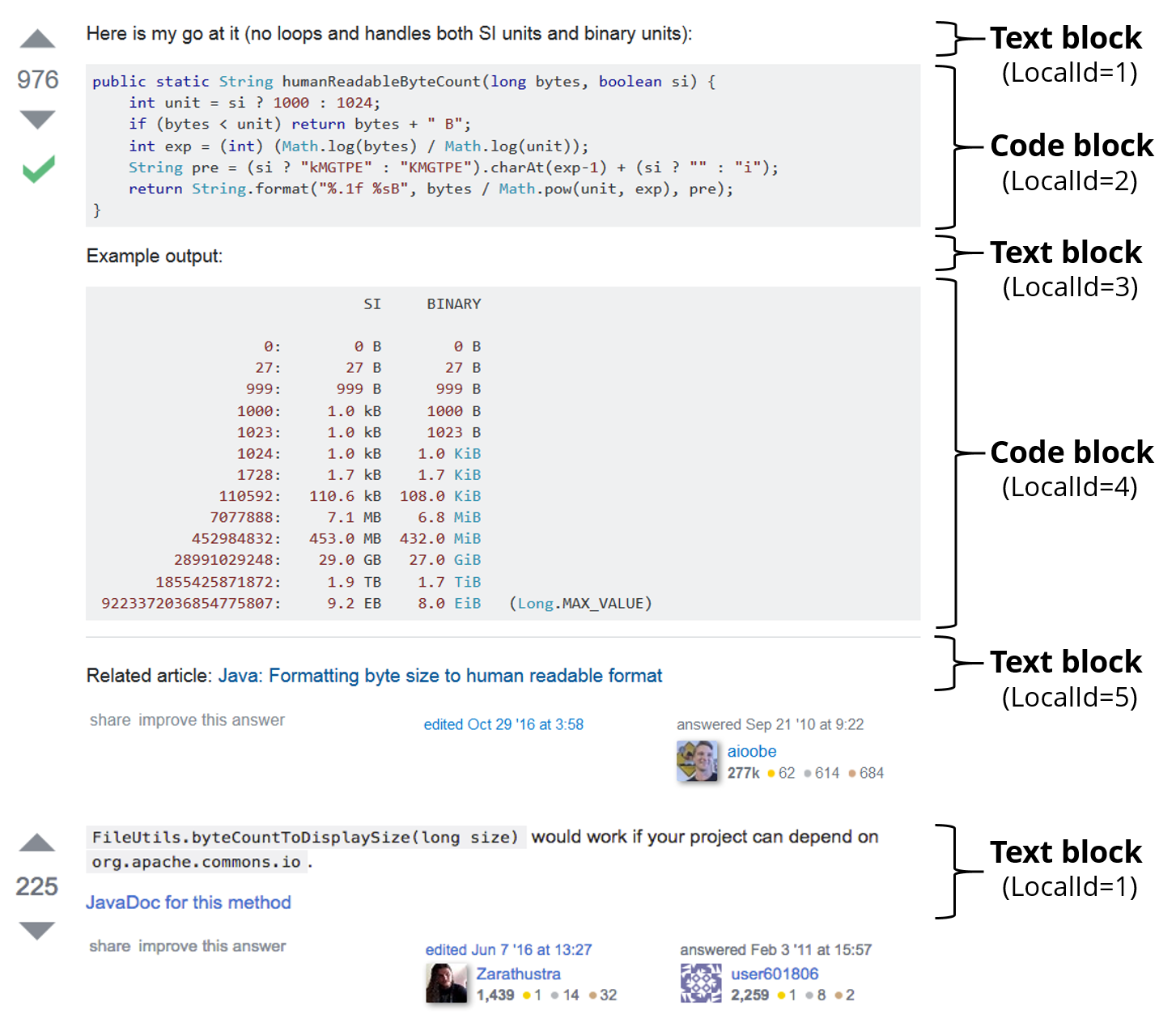} 
\caption{Exemplary Stack Overflow answers with code blocks (top, \href{https://stackoverflow.com/a/3758880}{3758880}) and with inline code (bottom, \href{https://stackoverflow.com/a/4888400}{4888400}). The \texttt{LocalId} represents the position in the post.}
\label{fig:so-postblocks-example}
\end{figure}

\begin{figure}[h!]
\centering
\includegraphics[width=1\columnwidth,  trim=0.0in 0.0in 0.0in 0.0in]{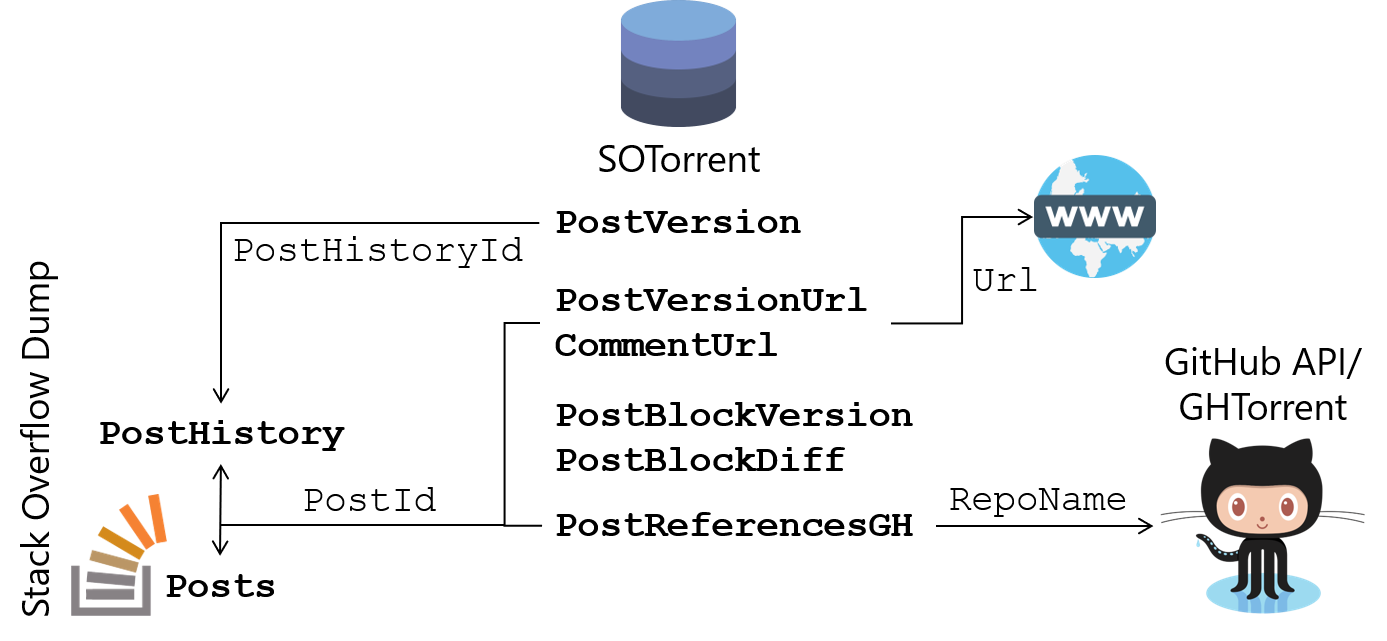} 
\caption{Connection of \emph{SOTorrent} tables to other resources.}
\label{fig:linking}
\end{figure}

\emph{SOTorrent} is an open dataset based on data from the official SO data dump~\cite{StackExchangeInc2017d} and the Google BigQuery GitHub (GH) dataset~\cite{GoogleCloudPlatform2018} that enables researchers to analyze the version history of SO posts at the level of individual text and code blocks (see Figure~\ref{fig:so-postblocks-example} for exemplary posts).
The official SO data dump~\cite{StackExchangeInc2017d} keeps track of different versions of entire posts, but does not contain information about differences between versions at a more fine-grained level.
In particular, extracting different versions of the same code snippet from the history of a post is challenging and required us to develop a complex strategy, involving the evaluation of 134 different string similarity metrics~\cite{BaltesDumaniOthers2018}.
Beside providing access to the version history, our dataset links SO posts to external resources in two ways: (1) by extracting linked URLs from text blocks of SO posts and from post comments and (2) by providing a table with links to SO posts found in the source code of open source GH projects.
This table can be used to connect \emph{SOTorrent} to GH datasets such as \emph{GHTorrent}~\cite{Gousios2013}. 
Analyses can be based on \emph{SOTorrent} alone or expanded to include data from other resources (see Figure~\ref{fig:linking}).
Questions that are, to the best of our knowledge, not sufficiently answered yet include:

\begin{itemize}
\item How are code snippets on SO maintained?
\item How many clones of code snippets exist inside SO?
\item How can we detect buggy versions of SO code snippets and find them in GH projects?
\item How frequently are code snippets copied from external sources into SO and then co-evolve there?
\item How do snippets copied from SO to GH co-evolve?
\item Does the evolution of SO code snippets follow patterns? 
\item Do these patterns differ between prog. languages?
\item Are the licenses of external sources compatible with SO's license (CC BY-SA 3.0)?
\item How many code blocks on SO do not contain source code (and are only used for markup)?
\item Can we reliably predict bug-fixing edits to code on SO?
\item Can we predict popularity of SO code snippets on GH?
\end{itemize}

\emph{SOTorrent} is available on Zenodo as a CSV database dump~\cite{BaltesDumani2018i} together with instructions on how to import the dataset. 
Moreover, the dataset is available as a Google BigQuery dataset~\cite{BaltesDumani2018j}, which allows to execute complex queries without the need to import the dataset (1 TB of queries per month are free).
We also published the source code of the software that we used to build~\cite{BaltesDumani2018c, Baltes2018b} and analyze~\cite{Baltes2018d, Baltes2018e} \emph{SOTorrent}.
More information about the dataset can be found in the corresponding research paper~\cite{BaltesDumaniOthers2018}.

\appendices

\section{Figures}

\section{Data Collection and Database Schema}

The \emph{SOTorrent} dataset contains all tables from the official Stack Overflow data dump.
However, that dump does only provide the version history at the level of whole posts as Markdown-formatted text.
To analyze how individual text or code blocks evolve, we needed to extract individual blocks from that content.
This extraction also enabled us to collect links to external sources from the identified text blocks.

In the SO dump, one version of a post corresponds to one row in the table \texttt{PostHistory}.
However, that table does not only document changes to the content of a post, but also changes to metadata such as tags or events such as closing of posts.
Since our goal was to analyze the evolution of SO posts at the level of whole posts and individual post blocks, we had to filter and process the available data.
First, we selected edits that changed the content of a SO post, identified by their \texttt{PostHistoryTypeId}~\cite{StackExchangeCommunityWiki20180227} ($2$: \textit{Initial Body}, $5$: \textit{Edit Body}, $8$: \textit{Rollback Body}).
We linked each filtered version to its predecessor and successor and stored it in table \texttt{PostVersion}.

The content of a post version is available as Markdown-formatted text.
We split the content of each version into text and code blocks and extracted the URLs from all text blocks using a regular expression (table \texttt{PostVersionUrl}).
We also extracted the URLs from all comments in the SO data dump (table \texttt{CommentUrl}).
Beside the extracted URLs, those tables provide information about the link type (e.g., bare, Markdown, or HTML), link position (top, middle, or end of post/comment), and certain URL components such as the root domain, query string, or fragment identifier (if present).
To reconstruct the version history of individual post blocks (table \texttt{PostBlockVersion}), we established a linear predecessor relationship between the post block versions using a string similarity metric that we selected after a thorough evaluation~\cite{BaltesDumaniOthers2018}.
For each post block version, we computed the line-based difference to its predecessor, which is available in table \texttt{PostBlockDiff}.

We also extracted the version history of question titles from table \texttt{PostHistory}.
Table \texttt{TitleVersion} links all title versions to their predecessors and successors and further provides the corresponding Levenshtein distances (columns \texttt{PredEditDistance} and \texttt{SuccEditDistance}).

One row in table \texttt{PostReferenceGH} represents one link from a file in a public GH repository to a post on SO.
To extract those references, we utilized Google BigQuery, which allows to execute SQL queries on various public datasets, including a dataset with all files in the default branch of GH projects~\cite{GoogleCloudPlatform2018}.
To find references to SO, we again applied a regular expression and mapped all extracted URLs to their corresponding sharing link (ending with \verb+/q/<id>+ for questions and \verb+/a/<id>+ for answers), storing that link together with information about the file and the repository in which the link was found in table \texttt{PostReferenceGH}.
We ignored other links referring to, e.g., users or tags on SO.

Version 2018-08-28 of the \emph{SOTorrent} dataset contains the version history of all 40,606,950 questions and answers in the official SO data dump published June 5, 2018~\cite{StackExchangeInc2017b}.
It contains 63,914,798 post versions, 122,673,430 text block versions, and 77,578,494 code block versions, ranging from the creation of the first post on July 31, 2008 until the last edit on June 3, 2018.
We extracted links to 11,775,659 distinct URLs from 20,518,181 different post block versions and 4,104,869 distinct URLs from 6,856,777 different comments.
Moreover, we identified 6,035,737 links to SO posts in 436,615 public GH repositories. 

Our project website~\footnote{\url{http://sotorrent.org}} lists all dataset versions and contains more information on the database layout, including the complete database schema.

\section{Data Sample}


To illustrate how researchers can use \emph{SOTorrent} to analyze the evolution of SO posts, we investigate one of the most popular Java answers on SO, which is depicted in Figure~\ref{fig:so-postblocks-example}.
It is the accepted answer for the question \emph{``How to convert byte size into human readable format in java?''}.\footnote{\url{https://stackoverflow.com/q/3758606}}
The following SQL queries are tested on a MySQL 5.7 database system with \emph{SOTorrent} 2018-08-28, but will also work with later versions.
Since some \emph{SOTorrent}-specific IDs (e.g., \texttt{PostVersionId} or \texttt{PostBlockVersionId}) may change between dataset versions, it is recommended to use the IDs from the official Stack Overflow data dump when possible (e.g., \texttt{PostId} or \texttt{PostHistoryId}), which are included as foreign keys and are stable across all dataset versions.
Exemplary BigQuery queries can be found in an earlier blog post about \emph{SOTorrent}~\footnote{\url{http://empirical-software.engineering/blog/sotorrent}}.
We start our investigation by retrieving all post block versions of the above-mentioned answer using its \texttt{PostId}:

\begin{sql}
SELECT PostHistoryId, PostBlockTypeId, LocalId, ...      # see Fig. 3 for all selected columns
FROM PostBlockVersion
WHERE PostId=3758880
ORDER BY PostHistoryId ASC, LocalId ASC;
\end{sql}

\begin{figure*}
\centering
\includegraphics[width=1\textwidth,  trim=0.0in 0.2in 0.0in 0.0in]{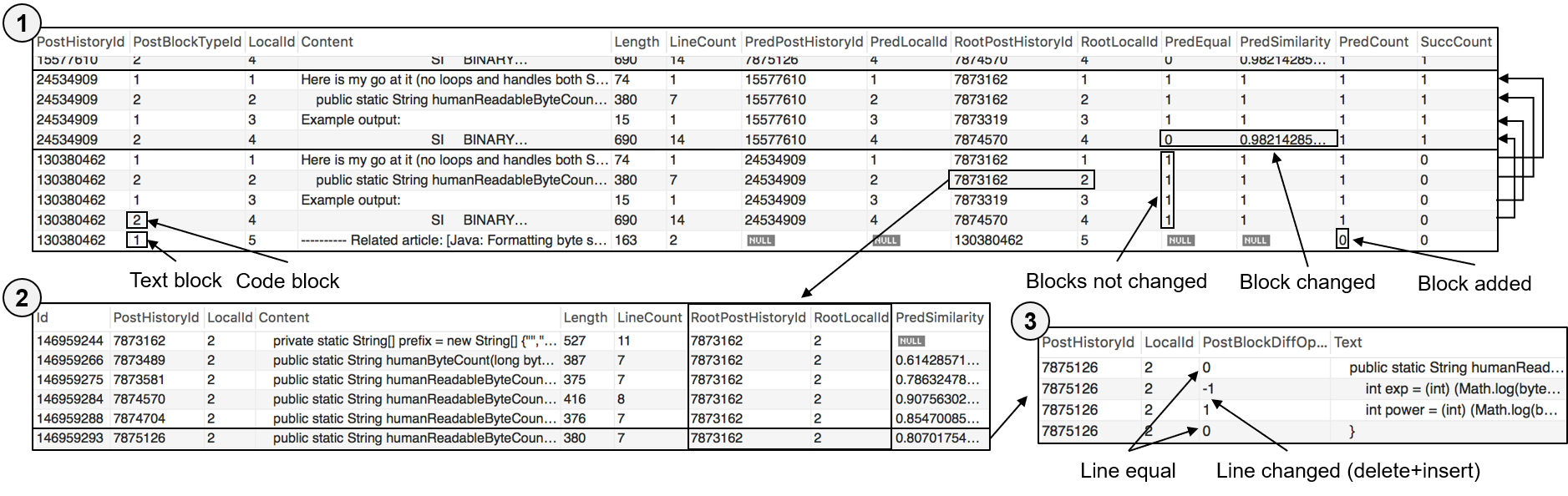}
\caption{Version data for Stack Overflow answer with ID \href{https://stackoverflow.com/a/3758880}{3758880} (truncated, based on tables \texttt{PostBlockVersion} and \texttt{PostBlockDiff}).}
\label{fig:example-tables}
\end{figure*}

Part (1) of Figure~\ref{fig:example-tables} shows the result of the above query for the two most recent post versions.
In our \emph{SOTorrent} paper~\cite{BaltesDumaniOthers2018}, we defined post block lifespans as chains of connected post block versions that are predecessors of each other.
Those chains can be easily retrieved from the database, because each post block version points to its \texttt{RootPostHistoryId} and \texttt{RootLocalId}.
Those columns uniquely identify the first post block version in the chain.
For part (2) of the figure, we only selected versions of the code snippet in the answer in which the content was actually modified (not all post blocks are modified in all post versions):

\begin{sql}
SELECT Id, PostHistoryId, LocalId, Content, Length, ...  # see Fig. 3 for all selected columns
FROM PostBlockVersion
WHERE RootPostHistoryId=7873162 AND RootLocalId=2
  AND (PredEqual IS NULL OR PredEqual = 0)
ORDER BY PostHistoryId ASC;
\end{sql}

To further see which lines of a code snippet were changed in the last edit, we can utilize table \texttt{PostBlockDiff}.
The result of the following query is shown in part (3) of the figure:

\begin{sql}
SELECT PostHistoryId, LocalId, PostBlockDiffOperationId, Text
FROM PostBlockDiff
WHERE PostHistoryId=7875126 AND LocalId=2;
\end{sql}

We can also use \emph{SOTorrent} to retrieve files on GitHub that reference this particular Stack Overflow post:

\begin{sql}
SELECT RepoName, Branch, Path, FileExt, Copies, PostId, SOUrl, GHUrl
FROM PostReferenceGH
WHERE PostId=3758880;
\end{sql}

\begin{figure*}
\centering
\includegraphics[width=0.9\textwidth,  trim=0.0in 0.1in 0.0in 0.1in]{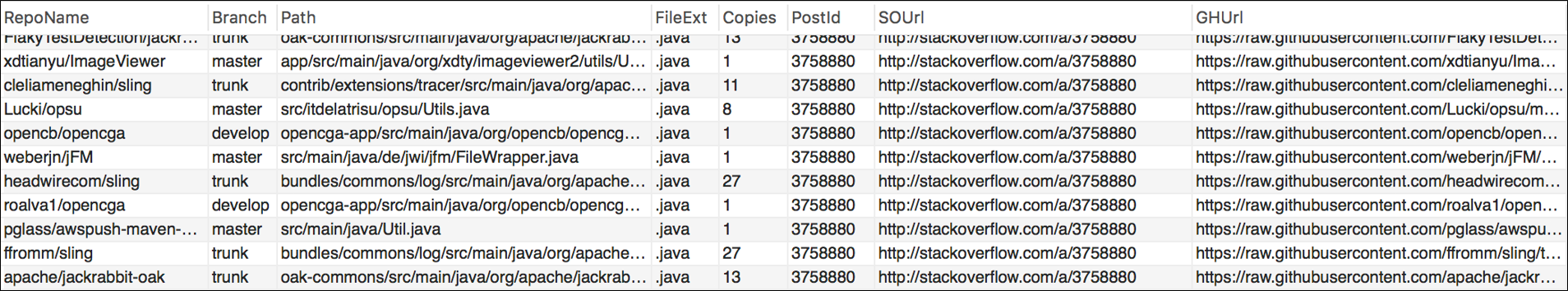}
\caption{GitHub references to Stack Overflow answer \href{https://stackoverflow.com/a/3758880}{3758880} (table \texttt{PostReferenceGH}, truncated).}
\label{fig:example-gh-references}
\end{figure*}

The result of this query is shown in Figure~\ref{fig:example-gh-references}.
To retrieve links from all text block versions of the post, we can use table \texttt{PostVersionUrl}:

\begin{sql}
SELECT *
FROM PostVersionUrl
WHERE PostId=3758880;
\end{sql}

In this case, only one post block version contains a link, which refers to a blog post with the same snippet (see Figure~\ref{fig:so-postblocks-example}).
To retrieve the title versions of the question that started the thread, we can use the following query:

\begin{sql}
SELECT *
FROM TitleVersion
WHERE PostId=(
  SELECT ParentId
  FROM Posts
  WHERE Id=3758880
);
\end{sql}

For this thread, however, the title has never been changed.
Retrieving all links from comments to this particular post is as simple as:

\begin{sql}
SELECT *
FROM CommentUrl
WHERE PostId=3758880;
\end{sql}

This query reveals one link in a comment, pointing to a class in the iOS API providing a functionality similar to the snippet in the post.

In our previous \emph{SOTorrent} paper~\cite{BaltesDumaniOthers2018}, we described a close relationship between post edits and comments.
To support a further investigation of this relationship, we wrote a blog post\footnote{\url{http://empirical-software.engineering/blog/sotorrent-edithistory}} showing how to create a new table \texttt{EditHistory}, which aggregates all title and body edits of Stack Overflow posts, together with post comments.
Using this table (and a helper table \texttt{Threads}), one can easily retrieve the edit and comment history of individual threads (see blog post for more details):

\begin{sql}
SELECT * FROM EditHistory
WHERE PostId IN (
	SELECT PostId FROM Threads WHERE ParentId = (
	  SELECT ParentID FROM Threads
	  WHERE PostId=3758880
	  # the question PostId 3758606 yields the same result
)) ORDER BY CreationDate;
\end{sql}

\section*{Acknowledgments}
The authors would like to thank Lorik Dumani for his help in evaluating different string similarity metrics for reconstructing the version history of Stack Overflow post blocks.

\bibliographystyle{IEEEtran}
\bibliography{literature}

\end{document}